\def\etal{{\it et al.}}
\def\ie{{\it i.e.}}

\def\~{{$\tilde{\phantom{a}}$}}

\documentclass [12pt] {article}
\usepackage{epsfig}
\usepackage{color}

\def\thebibliography#1{\section{References}\markboth
 {REFERENCES}{REFERENCES}\list
 {[\arabic{enumi}]}{\settowidth\labelwidth{[#1]}\leftmargin\labelwidth
 \advance\leftmargin\labelsep
 \usecounter{enumi}}
 \def\newblock{\hskip .11em plus .33em minus -.07em}
 \sloppy
 \sfcode`\.=1000\relax}
\def\upcite#1{\raise6pt\hbox{\scriptsize
\cite{#1}}}
\def\lsim{\mathrel {\vcenter {\baselineskip 0pt \kern 0pt
    \hbox{$<$} \kern 0pt \hbox{$\sim$} }}}
\def\gsim{\mathrel {\vcenter {\baselineskip 0pt \kern 0pt
    \hbox{$>$} \kern 0pt \hbox{$\sim$} }}}
\def\gtlt{\mathrel {\vcenter {\baselineskip 0pt \kern 0pt
    \hbox{$>$} \kern 0pt \hbox{$<$} }}}

\def\hline{\noalign{\hrule \vskip2pt}}

\def\|{\ifmmode\Vert\else \char`\|\fi}
\ifx\oldzeta\undefined                          
  \let\oldzeta=\zeta                            
  \def\zzeta{{\raise 2pt\hbox{$\oldzeta$}}}     
  \let\zeta=\zzeta                              
\fi

\ifx\oldchi\undefined                           
  \let\oldchi=\chi                              
  \def\cchi{{\raise 2pt\hbox{$\oldchi$}}}       
  \let\chi=\cchi                                
\fi

\def\frac#1#2{{#1 \over #2}}

\def\half{\ifinner {\scriptstyle {1 \over 2}}
   \else {1 \over 2} \fi}

\def\abs#1{\left\vert#1\right\vert}	

\def\simge{\mathrel{%
   \rlap{\raise 0.511ex \hbox{$>$}}{\lower 0.511ex \hbox{$\sim$}}}}
\def\simle{\mathrel{
   \rlap{\raise 0.511ex \hbox{$<$}}{\lower 0.511ex \hbox{$\sim$}}}}

\def\buildchar#1#2#3{{\null\!                   
   \mathop#1\limits^{#2}_{#3}                   
   \!\null}}                                    
\def\overcirc#1{\buildchar{#1}{\circ}{}}

\def\slashchar#1{\setbox0=\hbox{$#1$}           
   \dimen0=\wd0                                 
   \setbox1=\hbox{/} \dimen1=\wd1               
   \ifdim\dimen0>\dimen1                        
      \rlap{\hbox to \dimen0{\hfil/\hfil}}      
      #1                                        
   \else                                        
      \rlap{\hbox to \dimen1{\hfil$#1$\hfil}}   
      /                                         
   \fi}                                         %

\def\subrightarrow#1{
  \setbox0=\hbox{
    $\displaystyle\mathop{}
    \limits_{#1}$}
  \dimen0=\wd0
  \advance \dimen0 by .5em
  \mathrel{
    \mathop{\hbox to \dimen0{\rightarrowfill}}
       \limits_{#1}}}                           

\def\overlay#1#2{\ifmmode%
\setbox0=\hbox{$#1$}%
\setbox1=\hbox to\wd0{\hss$#2$\hss}\else%
\setbox0=\hbox{#1}%
\setbox1=\hbox to\wd0{\hss#2\hss}\fi%
#1\hskip-\wd0\box1 }

\def\pmb#1{\leavevmode\setbox0=\hbox{#1}%
\kern-.02em\copy0\kern-\wd0
\kern.04em\copy0\kern-\wd0
\kern-.02em\raise.04em\box0 }

\def\vereq#1#2{\lower3pt\vbox{\baselineskip1.5pt \lineskip1.5pt
\ialign{$\m@th#1\hfill##\hfil$\crcr#2\crcr\sim\crcr}}}

\def\tensor#1{\protect\@ontopof{#1}{\leftrightarrow}{1.15}\mathord{\box2}}
\def\overstar#1{\protect\@ontopof{#1}{\ast}{1.15}\mathord{\box2}}
\def\overdots#1{\protect\@ontopof{#1}{\cdots}{1.0}\mathord{\box2}}
\def\overcirc#1{\protect\@ontopof{#1}{\circ}{1.2}\mathord{\box2}}
\def\loarrow#1{\protect\@ontopof{#1}{\leftarrow}{1.15}\mathord{\box2}}
\def\roarrow#1{\protect\@ontopof{#1}{\rightarrow}{1.15}\mathord{\box2}}

\def\@ontopof#1#2#3{%
{\mathchoice
{\@@ontopof{#1}{#2}{#3}\displaystyle\scriptstyle}%
{\@@ontopof{#1}{#2}{#3}\textstyle\scriptstyle}%
{\@@ontopof{#1}{#2}{#3}\scriptstyle\scriptscriptstyle}%
{\@@ontopof{#1}{#2}{#3}\scriptscriptstyle\scriptscriptstyle}%
}%
}

\def\@@ontopof#1#2#3#4#5{%
\setbox0=\hbox{$#4#1$}%
\setbox1=\hbox{$#5#2$}%
\setbox2=\hbox{}\ht2=\ht0 \dp2=\dp0 %
\ifdim\wd0>\wd1 %
\setbox1=\hbox to\wd0{\hss\box1\hss}%
\mathord{\rlap{\raise#3\ht0\box1}\box0}%
\else   %
\setbox1=\hbox to.9\wd1{\hss\box1\hss}%
\setbox0=\hbox to\wd1{\hss$#4\relax#1$\hss}%
\mathord{\rlap{\copy0}\raise#3\ht0\box1}%
\fi
}%

\def\lambdabar{\protect\@lambdabar}
\def\@lambdabar{%
\relax
\bgroup
\def\@tempa{\hbox{\raise.73\ht0
\hbox to0pt{\kern.25\wd0\vrule width.5\wd0
height.1pt depth.1pt\hss}\box0}}%
\mathchoice{\setbox0\hbox{$\displaystyle\lambda$}\@tempa}%
{\setbox0\hbox{$\textstyle\lambda$}\@tempa}%
{\setbox0\hbox{$\scriptstyle\lambda$}\@tempa}%
{\setbox0\hbox{$\scriptscriptstyle\lambda$}\@tempa}%
\egroup
}

\def\corresponds{{\lower.2ex\hbox{=}}{\rm\kern-.75em^\triangle}}
\def\succsim{\succ\kern-.9em_\sim\kern.3em}
\def\precsim{\prec\kern-1em_\sim\kern.3em}
\def\slantfrac#1#2{\kern1em^{#1}\kern-.3em/\kern-.1em_{#2}}

\begin{document}

\begin{center}
{\Large\bf A Neutrino Horn Based on a Solenoid Lens}
\\

\medskip

Kirk T.~McDonald
\\
{\sl Joseph Henry Laboratories, Princeton University, Princeton, NJ 08544}
\\
(December 1, 2003)
\end{center}

\section{Problem}

This note considers variations on the theme of a solenoid magnet (\ie, a magnet
whose field has axial symmetry) as a
lens for charged particles.  A related problem has been posed in \cite{canon}.

Recall that if a device is to be a lens with optic axis along the $z$ axis
in a cylindrical coordinate system $(r,\phi,z)$,
then as particles leave the device they must have no azimuthal momentum,
$P_\phi = 0$, and their
radial momentum must be proportional to their radial coordinate,
$P_r \propto r$.  Special cases are (1) that all particles have $r = 0$ at
the exit of the device, which is a focal point; and (2) that all particles
have zero radial momentum.

\subsection{Particle Source Inside the Solenoid: A Neutrino ``Horn"}

A neutrino ``horn'' is a magnetic device whose goal is to focus
charged $\pi$ mesons that emerge from a target into a parallel beam,
so that when the pions decay, $\pi^\pm \to \mu^\pm \nu$, the
resulting neutrinos form a beam that has minimal angular 
divergence.\footnote{Because of the Jacobean peak in the two-body
decay kinematics of the pion, for some purposes it is favorable to use neutrinos
produced at a nonzero decay angle.  See, for example, \cite{offaxis}.}
Suppose the pions are produced at the origin, inside a solenoid magnet
of uniform field ${\bf B} = B \hat{\bf z}$ whose axis is the $z$ axis
and whose downstream face is at $z = L$.  Show that pions of momenta 
\begin{equation}
P  = {e B L \over (2 n + 1) \pi c }\, , \quad (n = 0, 1, 2, ....)
\label{p0}
\end{equation}
emerge from the magnet with their momenta parallel to the $z$ axis,
independent of the production angle $\theta$ (for $\theta \ll 1$).
In this case, the solenoid acts like an ideal thin lens of focal length $L$,
located at $z = L$.

Neutrinos from the forward decay of the resulting parallel beam of pions 
will have a quasi line spectrum with momenta proportional to those of
eq.~(\ref{p0}).  If the neutrinos are detected at a distance $l$ from
the source, that distance can be chosen so that the various peaks in
the neutrino spectrum all satisfy the condition for maximal probability 
of oscillation into another neutrino species prior to their detection.

\subsection{Particle Source Outside the Magnet}

Consider a point source of charged particles located at a distance $D$
from the entrance to solenoid magnet of length $L$ and field strength $B$, 
the source being on
the magnetic axis.  For what momenta $P$ are particles with angle $\theta
\ll 1$ with respect to the magnetic axis focused to a point
on axis beyond the exit of the magnet?

\bigskip

In both cases, the focusing effect is due to the fringe field of the magnet,
and not due to the uniform central field.  A simple model of this effect 
(impulse approximation) supposes
the magnetic ``kicks'' of the fringe field occur entirely in the entrance
and exit planes of the magnet.  Although this effect can be analyzed by
direct use of $F = ma$, it is helpful to consider the canonical (angular) momentum
of the particle in the magnetic field.  For this, you can use either
a Lagrangian formulation, or direction calculation via the Lorentz force law,
in which latter case first consider $d L_z / dt = d ({\bf r} \times {\bf P})_z / dt$.

\section{Solution}

Although this problem can be solved without explicit use of the canonical angular
momentum of a charged particle in a magnetic field, that concept offers an
elegant perspective.  Therefore, we first discuss canonical momenta in sec.~2.1,
and then comment on the paraxial approximation in sec.~2.2, and the impulse approximation
in sec.~2.3, before turning to the solutions for solenoid focusing
of particles produced outside, and inside, of the magnet in secs.~2.4 and 2.5. 
The possibly novel aspect of this note is the discussion in sec.~2.5.1 of a neutrino
horn based on solenoid focusing.

\subsection{Conservation of Canonical Angular Momentum}

The canonical momentum of a particle of charge $e$ and rest mass $m$ 
is (in rectangular coordinates and in Gaussian units)
\begin{equation}
{\bf p} = {\bf P} + {e {\bf A} \over c}\, ,
\label{s1}
\end{equation}
where ${\bf P} = \gamma m {\bf v} = m {\bf v} / \sqrt{1 - v^2 / c^2}$
 is the mechanical momentum of the particle,
{\bf A} is the vector potential of the magnetic field, 
and $c$ is the speed of light.  The canonical angular momentum is
\begin{equation}
{\bf l} = {\bf r} \times {\bf p},
\label{s2a}
\end{equation}
where {\bf r} is the position vector of the particle. 

One way to deduce the conserved quantities for the particle's motion is to
consider its Lagrangian or Hamiltonian.
If an electric field is present as well,
with electric potential $V$, the Lagrangian ${\cal L}$ of the particle can be
written \cite{Landau}
\begin{equation}
{\cal L} = - {m c^2 \over \gamma} + {e {\bf A} \cdot {\bf v} \over c} - e V,
\label{s1a}
\end{equation}
where ${\bf v} = d{\bf r} / dt$ is the particle's velocity.  The canonical momentum
associated with a rectangular coordinate $x_i$ is therefore $p_i = \partial {\cal L}
/ \partial \dot x_i$, leading to eq.~(\ref{s1}).
Then, the Hamiltonian ${\cal H}$ of the system is
\begin{equation}
{\cal H} = \sqrt{m^2 c^4 + \left( {\bf p} - {e {\bf A} \over c} \right)^2} + e V.
\label{s1b}
\end{equation}

If the external electromagnetic fields have azimuthal symmetry, then the potentials
$V$ and {\bf A} do also.  We consider a cylindrical coordinate system $(r,\phi,z)$
with the $z$ axis being the axis of symmetry of the fields.  Then both the
Lagrangian and the Hamiltonian have no azimuthal dependence,
\begin{equation}
{\partial {\cal L} \over \partial \phi} = {\partial {\cal H} \over \partial \phi} = 0,
\label{s1c}
\end{equation}
so the equations of motion (and the identities ${\bf r} = r \hat{\bf r} + z \hat{\bf z}$,
$\dot{\bf r} = {\bf v} = \dot r \hat{\bf r} + r \dot\phi \hat{\phi} + \dot z \hat{\bf z}$)
tell us that the canonical momentum $p_\phi$ is a constant of the motion
(even for time-dependent fields, so long as
they are azimuthally symmetric),\footnote{Note that the definition 
(\ref{s1d}) of the canonical momentum $p_\phi$ leads to the awkward result that
$p_\phi = r ({\bf p})_\phi$, where $({\bf p})_\phi$ is the $\phi$ component of the
canonical momentum vector {\bf p} of eq.~(\ref{s1}).} 
\begin{equation}
p_\phi = {\partial {\cal L} \over \partial \dot\phi} 
= r \left( \gamma m r \dot\phi + {e A_\phi \over c} \right)
= r ({\bf p})_\phi
= l_z.
\label{s1d}
\end{equation}
We also see that the canonical momentum $p_\phi$ can be interpreted as
the $z$ component of the canonical angular momentum (\ref{s2a}), so $l_z$ is
also a constant of the motion.

\bigskip

For completeness, we verify that $dl_z / dt = 0$ using the Lorentz force law,
\begin{equation}
{d {\bf P} \over d t} = e \left( {\bf E} + {{\bf v} \over c} \times {\bf B} \right)
= e \left( - \nabla V - {1 \over c} {\partial {\bf A} \over \partial t}
 + {{\bf v} \over c} \times (\nabla \times {\bf A}) \right).
\label{s1e}
\end{equation}
We begin with the ordinary angular momentum ${\bf L} = {\bf r} \times {\bf P}$,
and consider the $z$ component of its time derivative:
\begin{equation}
{d L_z \over d t} = {d ({\bf r} \times {\bf P})_z \over d t}
= \left({\bf r} \times {d {\bf P} \over d t} \right)_z
= r \left( {d {\bf P} \over d t} \right)_\phi.
\label{s1f}
\end{equation}
From eq.~(\ref{s1e}) we have, since $\partial V / \partial \phi 
= \partial A_r / \partial \phi = \partial A_z / \partial \phi = 0$,
\begin{eqnarray}
\left({d {\bf P} \over d t} \right)_\phi  
& = & - {e \over c} \left( {\partial A_\phi \over \partial t}
 + {\dot r \over r} {\partial (r A_\phi) \over \partial r} 
+ \dot z {\partial A_\phi \over \partial z} \right)
= - {e \over c r} \left( {\partial (r A_\phi) \over \partial t}
 + \dot r {\partial (r A_\phi) \over \partial r} 
+ \dot z {\partial (r A_\phi) \over \partial z} \right)
\nonumber \\
& = & - {e \over c r} {d (r A_\phi) \over d t}\, ,
\label{s1g}
\end{eqnarray}
where ${d \over dt}$ when applied to a field such as the vector potential {\bf A}
is the convective derivative associated with the moving particle.

Noting that ${\bf P} = \gamma m (\dot r \hat{\bf r} + r \dot\phi \hat\phi + \dot z \hat{\bf z})$
and $\dot{\hat{\bf r}} = \dot\phi \hat\phi$, 
we also find
\begin{equation}
\left({d {\bf P} \over d t} \right)_\phi  
= {d P_\phi \over d t} + \dot\phi P_r
= {d (\gamma m r \dot\phi) \over d t} + \gamma m \dot r \dot\phi
= {1 \over r} {d (\gamma m r^2 \dot\phi) \over dt}
= {1 \over r} {d (r P_\phi) \over dt}\, .
\label{s1h}
\end{equation}
Combining eqs.~(\ref{s1f})-(\ref{s1h}), we have
\begin{equation}
{d L_z \over d t} = {d (r P_\phi) \over dt}
= - {e \over c} {d (r A_\phi) \over d t}\, .
\label{s1i}
\end{equation}
Hence,
\begin{equation}
{d \over dt}\left[ r \left( P_\phi + {e \over c} A_\phi \right) \right]
= {d l_z \over d t} =  {d p_\phi \over d t} = 0,
\label{s1j}
\end{equation}
as found by the Lagrangian method as well.

\subsection{The Paraxial Approximation}

We now turn our attention to the question of lenslike character of a solenoid magnet
as a charged particle moves from a region of uniform field to zero field, or vice versa.

Inside a uniform solenoidal magnetic field ${\bf B} = B \hat{\bf z}$, the trajectory
of the particle is a helix (whose axis is in general at some 
nonzero radius $r_0$ from the magnetic axis).  
The radius $R$ of the helix can be obtained from ${\bf F} = M{\bf a}
= e {\bf v} / c \times {\bf B}$
using the relativistic mass $M = \gamma m$.  The projection of the motion onto
a plane perpendicular to the magnetic axis is a circle of radius $R$ and the
projected velocity is $v_\perp$.  Hence,
\begin{equation}
{\gamma m v_\perp^2 \over R} = e {v_\perp \over c} B,
\label{s4}
\end{equation}
so that
\begin{equation}
R = {c P_\perp \over e B}\, ,
\label{s5}
\end{equation}
where $P_\perp = \gamma m v_\perp$ is the transverse momentum of the particle.
For a particle whose average velocity is in the $+ z$ direction, 
the sense of rotation around the helix is in the $-\hat\phi$ direction (Lenz' law).
The angular frequency $\omega$ of the rotation (called the Larmor or cyclotron
frequency) also follows from eq.~(\ref{s4}):
\begin{equation}
\omega = {v_\perp \over R} = {e B \over \gamma m c}\, .
\label{s6}
\end{equation}

If the solenoid magnet has length $L$, then the time $t$ required for the
particle to traverse the magnet is given by
\begin{equation}
t = {L \over v_z} = {L \over P_z / \gamma m}
= {\gamma m L \over P \cos\theta}\, ,
\label{s7}
\end{equation}
where $\theta$ is the production angle of the particle with respect
to the $z$ axis.  Hence, the trajectory of the particle rotates about the axis
of the helix by azimuthal angle $\phi_h$ as the particle traverses the
magnet, where
\begin{equation}
\phi_h = \omega\, t = {e B  \over \gamma m c} \cdot {\gamma m L \over P \cos\theta}
= {e B L \over c P \cos\theta}\, .
\label{s8}
\end{equation}
There is a unique value for $\phi_h$ 
only for small production angles ($\theta \ll 1$), which is called the
{\bf paraxial} regime:
\begin{equation}
\phi_h \approx {e B \over c P} L = {L \over \lambdabar},
\qquad (\mbox{paraxial approximation},\ \theta \ll 1),
\label{s9}
\end{equation}
where we define the (reduced) Larmor wavelength of the particle's motion to be 
\begin{equation}
\lambdabar \equiv {c P \over e B}\, .
\label{s10}
\end{equation}

In the paraxial approximation the magnetic force that bends the particle's
trajectory into a helix is a weak effect, in that it depends on the product
of the small transverse velocity $v_\perp = v \sin\theta \ll v$ and the axial field $B$.

\subsection{The Impulse Approximation}

As the trajectory crosses the fringe field of the solenoid, the axial field
drops rapidly from $B$ to zero (or rises rapidly from zero to $B$).  
In this region there must be a radial
component to the magnetic field, according to the Maxwell equation
\begin{equation}
0 = \nabla \cdot {\bf B} 
= {1 \over r} {\partial (r B_r) \over r} + {\partial B_z \over \partial z}\, ,
\label{s51}
\end{equation}
so that
\begin{equation}
B_r \approx - {r \over 2} {\partial B_z \over \partial z}
\label{s52}
\end{equation}
(as also readily deduced by applying Gauss' law to a ``pillbox'' of radius $r$
and thickness $dz$).  Although the radial component $B_r$ 
of the magnetic field is small, it couples to
the large axial velocity $v_z$ to give a force $F_\phi = d P_\phi / dt$
in the azimuthal direction that is not negligible.  We can write
\begin{equation}
{d P_\phi \over dz} = {1 \over v_z} {d P_\phi \over dt}
= {1 \over v_z} {e v_z B_r \over c}
\approx - {e r \over 2 c} {\partial B_z \over \partial z}
\label{s53}
\end{equation}
Hence, the change $\Delta P_\phi$ in the azimuthal momentum of the particle as it
crosses the fringe field is
\begin{equation}
\Delta P_\phi \approx - {e r \Delta B_z \over 2 c} = {e r \Delta B \over 2 c}\, ,
\label{s54}
\end{equation}
since $\Delta B_z = - B$ at the axial field falls from $B$ to zero.

The {\bf impulse approximation} is that during the particle's passage through the
fringe field we can neglect the change in its momentum due to coupling with the
axial magnetic field.  We only consider the azimuthal kick (\ref{s54}).  Thus
\begin{equation}
P_{r,\rm out} = P_{r,\rm in}, \qquad
P_{\phi,\rm out} = P_{\phi,\rm in} + {e r B \over 2 c}\, , \qquad
P_{z,\rm out} = P_{z,\rm in}
\qquad \mbox{(impulse\ approximation)}.
\label{s55}
\end{equation}
Furthermore, we neglect the change in the transverse coordinates of the particle
as it passes through the fringe field.
\begin{equation}
r_{\rm out} = r_{\rm in}, \qquad
\phi_{\rm out} = \phi_{\rm in}
\qquad \mbox{(impulse\ approximation)}.
\label{s56}
\end{equation}

We can connect the impulse approximation with conservation of canonical angular
momentum by noting that a solenoid magnet with (uniform)
field ${\bf B} = B \hat{\bf z}$ has vector potential 
\begin{equation}
{\bf A} = A_\phi \hat\phi = {r B \over 2} \hat\phi.
\label{s57}
\end{equation}
To see this, recall that ${\bf B} = \nabla \times {\bf A}$
implies that the integral of the vector potential around a loop
is equal to the magnetic flux through the loop;
hence, $2 \pi r A_\phi = \pi r^2 B$.

The $z$ component of the canonical angular momentum (which is equal to the
azimuthal component of the canonical momentum $p_\phi$), 
\begin{equation}
l_z = p_\phi = r(P_\phi + e A_\phi / c)
= r(P_\phi + e r B / 2 c),
\label{s58}
\end{equation}
is a constant of the motion for a particle in
a solenoid magnet.   Hence, we see that the simplified impulse approximation
that $r_{\rm out} = r_{\rm in}$ plus conservation of canonical angular momentum
implies the form (\ref{s55}).

Additionally, we note that particles which are created
on the magnetic axis have $l_z = 0$, whether they are created inside or outside
the magnetic field.  As a consequence, whenever such a particle
is outside the magnetic field region it has $P_\phi = 0$.  If it has passed through
a region of solenoidal magnetic field, the azimuthal kicks at the entrance and
exit cancel exactly.  This results does not depend on the impulse approximation,
as it is deduced directly from conservation of canonical angular momentum.

\subsection{Particle Source Outside the Magnet}

\begin{figure}[htp]  
\begin{center}
\vspace{0.1in}
\includegraphics*[width=4in]{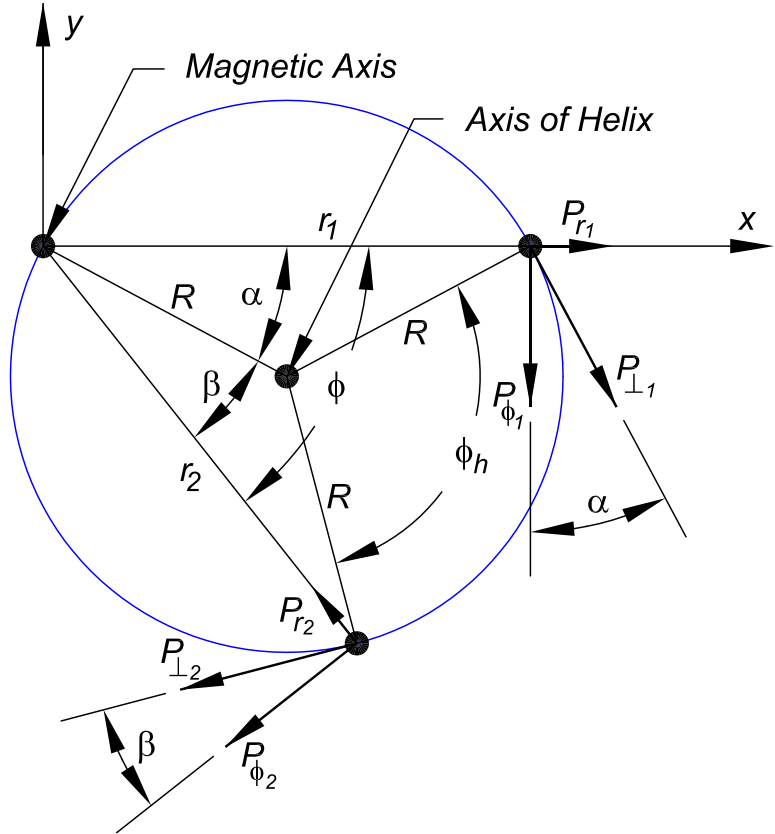}
\parbox{5.5in} 
{\caption[ Short caption for table of contents ]
{\label{fig3} Geometry of the helical trajectory of a particle of total 
momentum $P$ that enters
a solenoid magnet at $(r,\phi,z) = (r_1 = d_1\theta_1,0,0)$ with radial
momentum $P_{r_1} = P \theta_1$.  The fringe field at the entrance
of the solenoid gives the particle an azimuthal kick resulting in
momentum $P_{\phi_1} = - e B r_1 / 2 c$, where the magnetic field
is ${\bf B} = B \hat{\bf z}$ inside the solenoid.  
 The helix has radius $R = c P_\perp / e B$.  At the exit of
the solenoid the particle is at $(r_2,\phi,L)$ where $\phi = - e B L /
2 c P = \phi_h / 2$; the azimuthal rotation of the particle's
trajectory about the magnetic axis is one half that about the axis of
the helix.
}}
\end{center}
\end{figure}

We consider a solenoid magnet whose axis is the $z$ axis with
field ${\bf B} = B \hat{\bf z}$ for $0 < z < L$.  A particle
of momentum $P$ and charge $e$ is emitted at polar angle $\theta_1 \ll 1$
from a (point) source at $(x,y,z) = (0,0,-d_1)$, and so arrives
at the entrance of the magnet with spatial coordinates
$(r,\phi,z) \approx (r_1 = d_1 \theta_1,0,0)$ in the small angle
(paraxial) approximation, and with momentum $(P_r,P_\phi,P_z)
\approx (P_{r_1},0,P)$, where
\begin{equation}
P_{r_1} = P \theta_1.
\label{s100}
\end{equation}
The projection of the particle's
trajectory onto the $x$-$y$ plane is shown in Fig.~\ref{fig3}.

The fringe field at the entrance
of the solenoid gives the particle an azimuthal kick resulting in
momentum 
\begin{equation}
P_{\phi_1} = - {e B r_1 \over 2 c}
= - {e B d_1 \theta_1 \over 2 c}\, ,
\label{s101}
\end{equation} 
according to eq.~(\ref{s55}),
where the magnetic field is ${\bf B} = B \hat{\bf z}$ inside the solenoid. 
The transverse momentum $P_\perp$ of the particle inside the magnet is
therefore
\begin{equation}
P_\perp = \sqrt{P^2_{r_1} + P^2_{\phi_1}} 
= {e B r_1 \over 2 c} \sqrt{1 + \left( {2 c P \over e B d_1} \right)^2}
= {e B r_1 \over 2 c} \sqrt{1 + \left( {2 \lambdabar \over d_1} \right)^2}
=  {e B R \over c}\, ,
\label{s102}
\end{equation}  
where $R$ is the radius of the helical trajectory of the particle
inside the solenoid, recalling eq.~(\ref{s5}).  We also can write
\begin{equation}
r_1 = 2 R \cos\alpha,
\label{s103}
\end{equation}
where the angle $\alpha$, shown in Fig.~\ref{fig3}, is related by
\begin{equation}
\tan\alpha = {P_{r_1} \over \abs{P_{\phi_1}}}
= {2 c P \over e B d_1}
= {2 \lambdabar \over d_1}\, ,
\label{s104}
\end{equation}
which is independent of the production angle $\theta_1$ in
the paraxial approximation.

As the particle traverses length $L$ of the solenoid, its
trajectory rotates by azimuthal angle
\begin{equation}
\phi_h = - {e B L \over c P} = - {L \over \lambdabar}
\label{s105}
\end{equation}
about the axis of the helix.  At the exit of
the solenoid the particle is at $(r_2,\phi,L)$ in cylindrical
coordinates centered on the axis of the magnet (rather than on
the axis of the helix), as shown in Fig.~\ref{fig3}.
By the well-known geometrical relation
that the angle subtended by an arc on a circle as viewed from
another point on that circle is one half the angle subtended by
that arc from the center of the circle, we have that\footnote{
The geometrical relation (\ref{s106}) has the consequence that
in a frame that rotates about the magnetic axis
at half the Larmor frequency (\ref{s6}), the particle's trajectory
is simple harmonic motion in a plane that contains the magnetic
axis \cite{Kim}.   However, we do not pursue this insight here.}
\begin{equation}
\phi = {\phi_h \over 2} = - {e B L \over 2 c P}
= - {L \over 2 \lambdabar}\, .
\label{s106}
\end{equation}
The radial coordinate of the particle at the exit of the solenoid is
\begin{equation}
r_2 = 2 R \cos\beta,
\label{s107}
\end{equation}
where angle $\beta$ is given by
\begin{equation}
\beta = \abs{\phi} - \alpha
= {L \over 2 \lambdabar} - \tan^{-1} \left( {2 \lambdabar \over d_1} \right).
\label{s108}
\end{equation}

When the particle is at the exit of the solenoid, but still inside it,
the transverse momentum vector ${\bf P}_{\perp_2}$ makes angle $\beta$
to the unit vector $\hat\phi$, as shown in Fig.~\ref{fig3}.  The
radial momentum of the particle $P_{r_2}$ at the exit of the magnet
is therefore
\begin{equation}
P_{r_2} = - P_\perp \sin\beta
= - P_\perp {r_2 \over 2 R} \tan\beta
= - {e B r_2 \over 2 c} {\tan{L \over 2 \lambdabar} - {2 \lambdabar \over d_1}
 \over 1 + {2 \lambdabar \over d_1} \tan{L \over 2 \lambdabar} }\, ,
\label{s109}
\end{equation}
using eqs.~(\ref{s102}) and (\ref{s107}),
while the azimuthal component $P_{\phi_2}$ obeys
\begin{equation}
P_{\phi_2} = - P_\perp \cos\beta
= - {e B r_2 \over 2 c}\, .
\label{s110}
\end{equation}

As the particle exits the magnet, the radial component of its
transverse momentum remains at the value of eq.~(\ref{s109}) in
the impulse approximation, while the azimuthal component increases
by $e B r_2 / 2 c$ over the value of eq.~(\ref{s110}) 
and hence vanishes, as expected since the canonical
angular momentum is zero.

Once the particle has exited the magnet its transverse momentum
is purely radial, with a value proportional to the radial
coordinate $r_2$ at the exit of the magnet.  This is lens-like
behavior, in that the particle will then cross the magnetic
axis at distance $d_2$ from the exit of the magnet, where
\begin{equation}
{r_2 \over d_2} = \theta_2 = {P_{r_2} \over P}\, .
\label{s111}
\end{equation}
and so
\begin{equation}
d_2 = {2 c P \over e B \tan\beta}
= 2 \lambdabar {1 + {2 \lambdabar \over d_1} \tan{L \over 2 \lambdabar}
 \over \tan{L \over 2 \lambdabar} - {2 \lambdabar \over d_1} }
= {f d_1 \over d_1 - f}  
\left( 1 + {2 \lambdabar \over d_1} \tan{L \over 2 \lambdabar} \right)\, , 
\label{s112}
\end{equation}
where
\begin{equation}
f = {2 \lambdabar \over \tan{L \over 2 \lambdabar}}\, .
\label{s113}
\end{equation}
When distance $d_2$ is positive the solenoid acts as a (thick) focusing lens.

For the special cases of point-to-parallel focusing ($d_2 \to \infty$)
and parallel-to-point focusing ($d_1 \to \infty$), the solenoid magnet
has focal length $f$ given by eq.~(\ref{s113}).

If $(2 \lambdabar / d_1) \tan(L / 2 \lambdabar) \ll 1$ then the object 
distance $d_1$ and the image distance $d_2$ obey the lens formula
\begin{equation}
{1 \over d_1} + {1 \over d_2} = {1 \over f}
\qquad \left(  \tan{L \over 2 \lambdabar} \ll {d_1 \over 2 \lambdabar} \right).
\label{s114}
\end{equation}

If in addition the length $L$ of the
solenoid is small compared to the Larmor wavelength $\lambdabar$
the solenoid can be called a thin lens, for which
\begin{equation}
f = {4 \lambdabar^2 \over L}
\qquad (\mbox{thin\ lens}: L \ll \lambdabar,\ L \ll d_1).
\label{s115}
\end{equation}
This weakly focusing limit is, however, seldom achieved in practical 
applications of solenoid magnets as focusing elements.

The results (\ref{s112})-(\ref{s115}) for thick- and thin-lens
focusing can be utilized in a transfer-matrix description of
particle transport through magnetic systems \cite{Weidemann}.

\subsection{Particle Source Inside the Magnet}

The case of a source of particles inside the solenoid magnet, say
at $z = 0$, can be treated as a special case of the analysis in
sec.~2.4 in which $d_1 = r_1 = 0$.  The angle $\alpha$ shown in
Fig.~\ref{fig3} is $\pi / 2$ in this case, so that angle $\beta$ is
\begin{equation}
\beta = {L \over 2 \lambdabar} - {\pi \over 2}\, .
\label{s116}
\end{equation}
The radial coordinate of the particle at the exit of the magnet is
\begin{equation}
r_2 = 2 R \cos\beta
 = 2 R \sin{L \over 2 \lambdabar}\, ,
\label{s117}
\end{equation}
and the image distance $d_2$ follows from eq.~(\ref{s112}) as
\begin{equation}
d_2 = {2 c P \over e B \tan\beta}
= - 2 \lambdabar \tan{L \over 2 \lambdabar}\, .
\label{s118}
\end{equation}
The radial momentum at the exit of the magnet is
\begin{equation}
P_{r_2} = - P_\perp \sin\beta
= - {e B R \over c} {r_2 \over 2 R} \tan\beta
= {e B r_2 \over 2 c} \cot{L \over 2 \lambdabar}\, .
\label{s119}
\end{equation}
according to eqs.~(\ref{s109}) and (\ref{s116}).

This is
lens-like behavior ($P_{r_2} \propto r_2$) for any length $L$
of the solenoid, with $L = n \pi \lambdabar$
being the boundary between focusing and defocusing.

For the special case that $L = 2 n \pi \lambdabar$ we have $d_2 = r_2 = 0$,
corresponding to an image of the source occuring at the exit of the magnet.

Of particular interest here is the special case that 
$L = (2 n + 1) \pi \lambdabar$, for which $d_2 = \infty$, $P_{r_2} = 0$,
and we have point-to-parallel focusing.
From Fig.~\ref{fig3} and eq.~(\ref{s119}) we see that the condition
for point-to-parallel focusing of a source inside the solenoid is that
the particle has completed an odd number of half turns on its helical
trajectory when it
reaches the end of the solenoid.  In this case we can say that the focal
length of the solenoid lens is just the length $L$,
\begin{equation}
f = L = (2 n + 1) \pi {c P \over e B}
\qquad \mbox{(point-to-parallel\ focus,\ source\ inside\ solenoid)}.
\label{s120}
\end{equation}

\subsubsection{Neutrino Horn: Point-to-Parallel Focus, 
$L = (2 n + 1) \pi c P / e B$}

\begin{figure}[htp]  
\begin{center}
\vspace{0.1in}
\includegraphics*[width=6.5in]{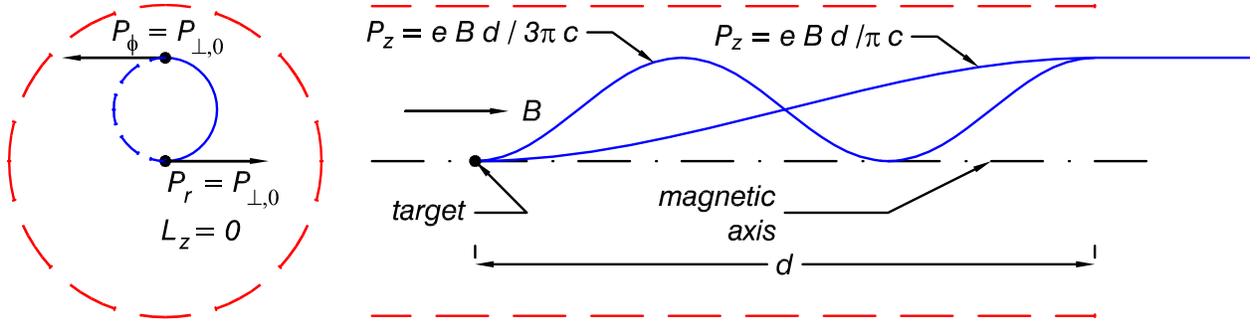}
\parbox{5.5in} 
{\caption[ Short caption for table of contents ]
{\label{fig2} Concept of a neutrino horn based on solenoid focusing.
The pion production target is inside the uniform field region of the
solenoid.  The focusing effects of the fringe field at the exit of
the magnet (at distance $L$ from the target) act as ideal thin lens
of focal length $L$ for a discrete set of particle momenta, given in
eq.~(\ref{s121}).
}}
\end{center}
\end{figure}

A solenoid magnet provides point-to-parallel focusing for particles
produced inside the magnet, on its axis, with a discrete set of
momenta $P_n$ given by
\begin{equation}
P_n  = {P_0 \over 2 n + 1}\, , \quad (n = 0, 1, 2, ....)
\qquad \mbox{where} \qquad
P_0  = {e  B L \over  \pi c}\, .
\label{s121}
\end{equation}
Particles with other momenta are not brought into parallelism, so that
a ``beam'' formed by drifting particles that emerge from the solenoid
will be quasimonochromatic with the sequence of momenta given in eq.~(\ref{s121}).
Figure~\ref{fig2} illustrates trajectories for particles of momenta $P_0$ and
$3 P_0$ in a solenoid magnet.

Such a sequence of momenta occurs in the phenomenon of neutrino oscillations over
a flight path $l$.  As is well, known, in the approximation of pure two-neutrino
mixing, the probability that neutrino type (mass eigenstate) 
$i$ of energy $E = P$ appears are neutrino
type $j$ after traversing distance $l$ is given by
\begin{equation}
\mbox{Prob}(i \to j) \propto \sin^2 {\Delta M^2_{ij} l \over 2 E}\, ,
\label{s122}
\end{equation}
where $\Delta M_{ij} = M_i - M_j$ is the difference in the masses of the
two neutrino types.  Hence, for a fixed drift distance $l$, the probability
of neutrino type $i$ appearing as type $j$ is maximal for the sequence of
neutrino momenta (energy)
\begin{equation}
P_n  = {P_0 \over 2 n + 1}\, , \quad (n = 0, 1, 2, ....)
\qquad \mbox{where} \qquad
P_0  = {\Delta M^2_{ij} l \over \pi }\, .
\label{s123}
\end{equation}
Thus a solenoid magnet could be very useful in preparing a neutrino beam
with a sequence of momenta such that all oscillation effects are maximal.
The potential advantage of such a beam for the study of CP violation in
neutrino oscillations has been pointed out by Marciano \cite{Marciano},
and elaborated upon in \cite{Diwan}.

Of course, neutrinos are neutral, so that a solenoid magnet cannot directly
affect their trajectories.  Rather, the solenoid magnet would be used to
focus $\pi^\pm$ particles that are produced in the interaction of a proton
beam with a nuclear target that is placed on the axis inside the magnet.
The length $l$ of the magnet should be short enough that most pions of
interest exit the magnet before decaying into neutrinos, according to
\begin{equation}
\pi^+ \to \mu^+ \nu_\mu, \qquad \pi^- \to \mu^- \bar\nu_\mu.
\label{s124}
\end{equation}
Because of the low ``Q'' value of this decay, the direction of the
neutrinos is very close to that of the pions, provided that latter have
energies greater than a few hundred MeV.  The forward-going neutrinos
carry about 4/9 of their parent pion momentum, so the solenoid system
should be chosen with a momentum $P_{0,\pi}$ equal to 9/4 of the highest
desired neutrino momentum at which the oscillation probability is
maximal, \ie,
\begin{equation}
P_{0,\pi} \approx {9 \over 4} P_{0,\nu}.
\label{s125}
\end{equation}

As implied by eq.~(\ref{s124}), the solenoid-focused beam would contain both
muon neutrinos and muon antineutrinos, in roughly equal numbers.  This has
the advantage to studies could be made simultaneously with both
neutrino and antineutrino beams.  However, for the study of CP violation
it would be necessary to identify whether each interactions was due to a
neutrino or an antineutrino.  This identification must be provided by
the detector in which the neutrino interacts. If the
neutrinos oscillate into electron neutrinos or antineutrinos before they
interact in a the detector, the latter must distinguish showers of
electrons from positrons.  This difficult experimental challenge can
likely only be met by a magnetized liquid argon detector
\cite{lanndd,microlanndd_cern,microlanndd_bnl}.

When studying the oscillation of muon neutrinos into electron neutrinos,
the presence of electron neutrinos in the beam constitutes the limiting
background.  Electron neutrinos are present in the beam due to the
3-body decay of the muons from pion decay:
\begin{equation}
\pi^+ \to \mu^+ \nu_\mu, \quad \mu^+ \to e^+ \nu_e \bar\nu_\mu,
\qquad \qquad
\pi^- \to \mu^- \bar\nu_\mu, \quad \mu^- \to e^- \bar\nu_e \nu_\mu.
\end{equation}
The background of electron neutrinos, compared to the flux of muon
neutrinos at a particular energy, is suppressed when the beam
contains only a narrow range of momenta of the parent pions.  This
occurs because the muon neutrinos from the pion decay then have
typically higher momentum that the electron neutrinos from the
related muon decay.  Hence, the solenoid-focused neutrino beam,
with its quasi line spectrum of energies will have lower
electron neutrino content, at least for highest-energy neutrino
``lines'', compared to a wide-band neutrino beam.

A final advantage of the solenoid-focused beam is that the magnetic
elements are farther removed transversely from the pion production
target, and so can be made more radiation resistant to intense
proton fluxes than is the case for more conventional toroid-focused
neutrino ``horns".  Further, the relatively open geometry of the
solenoid lens will permit use of liquid metal target, as needed if
the proton beam has several megawatts of power \cite{mercury}.

\bigskip

The author thanks Ron Davidson for the demonstration that conservation of the canonical
momentum $p_\phi$ follows from the Lorentz force law.

\end{document}